\documentclass[fleqn,usenatbib]{mnras}

\usepackage[T1]{fontenc}
\usepackage{ae,aecompl}

\usepackage{graphicx}
\usepackage{times}
\usepackage{array}
\usepackage{stfloats}
\usepackage{amsmath}
\usepackage{xspace}
\usepackage{txfonts}
\usepackage{ulem}

\graphicspath{{./images/}}

\newcommand{\ramses}{\textsc{RAMSES}\xspace}

\newcommand{\Planck}{\textit{Planck}\xspace}
\newcommand{\virg}[1]{`{#1}'}
\newcommand{\hMsol}{h^{-1}\,{\rm M_\odot}}

\newcommand{\hMpc}{h^{-1}\,{\rm Mpc}}
\newcommand{\hkpc}{h^{-1}\,{\rm kpc}}
\newcommand{\kms}{{\rm km\, s^{-1}}}

\newcommand{\enrico}[1]{{#1}}

\newcommand{\ie}{{i.e.~}}
\newcommand{\eg}{{e.g.~}}

\newcommand{\HI}{\textrm{H}\,\textsc{I}\xspace}
\newcommand{\HII}{\textrm{H}\,\textsc{II}\xspace}
\newcommand{\HeI}{\textrm{He}\,\textsc{I}\xspace}
\newcommand{\HeII}{\textrm{He}\,\textsc{II}\xspace}
\newcommand{\HeIII}{\textrm{He}\,\textsc{III}\xspace}
\newcommand{\afuv}{\alpha_{\rm FUV}}
\newcommand{\aeuv}{\alpha_{\rm EUV}}
\newcommand{\MAB}{M_{\rm AB}}   
\newcommand{\Mh}{M_{\rm h}}
\newcommand{\sAB}{\sigma_{\rm AB}}
\newcommand{\MH}{MH\xspace}
\newcommand{\CCPone}{CCP1\xspace}
\newcommand{\CCPtwo}{CCP2\xspace}
\newcommand{\Tzero}{T_{\rm 0}}
\newcommand{\Lya}{Ly$\alpha$\xspace}
\newcommand{\taueff}{\tau_{\rm eff}}
\newcommand{\mfp}{\lambda_{\rm mfp}}
\newcommand{\fesc}{f_{\rm esc}}
\newcommand{\citenp}[1]{\citeauthor{#1} \citeyear{#1}}  

\title[Quasar-only Reionization]{The Goldilocks problem of the quasar contribution to reionization}

\author[E. Garaldi et al.]{
Enrico Garaldi,$^{1}$\thanks{egaraldi@uni-bonn.de}\thanks{Member of the International Max Planck Research School (IMPRS) for Astronomy and Astrophysics at the Universities of Bonn and Cologne}
Michele Compostella,$^{2,3}$
and Cristiano Porciani$^{1}$
\\
$^{1}$Argelander Institut f\"ur Astronomie der Universit\"at Bonn, Auf dem H\"ugel 71, 53121 Bonn, Germany\\
$^{2}$Max Planck Institute for Astrophysics, Karl-Schwarzschild Stra{\ss}e 1, 85741 Garching, Germany\\
$^{3}$Max Planck Computing and Data Facility, Gie{\ss}enbachstra{\ss}e 2, 85741 Garching, Germany}

\date{Accepted XXX. Received YYY; in original form ZZZ}

\pubyear{2018}

\begin{document}
\label{firstpage}
\pagerange{\pageref{firstpage}--\pageref{lastpage}}
\maketitle

\begin{abstract}
The detection of an unexpectedly large number of
faint candidate quasars (QSO) at $z \gtrsim 4$ has motivated Madau and Haardt to investigate a reionization scenario in which active galactic nuclei dominate the photon budget at all times.
Their analytical study reveals that this picture is in 
agreement with the evolution of the \HI volume fraction in
the intergalactic medium (IGM)
and the optical depth to Thomson scattering of the cosmic microwave background.
We employ a suite of hydrodynamical simulations post-processed with a radiative transfer code to further investigate the properties of such a reionization history. 
In particular, we generate synthetic absorption-line spectra for \HI and \HeII that we compare with observational data. Although we confirm the analytical results mentioned above, we also find that
the evolution of the IGM temperature and of the \HeII optical depth 
are at odds with current observational constraints. 
Nevertheless, the QSO-dominated scenario presents the attractive feature that
it naturally generates an inhomogeneous IGM and
thus produces an extended tail at high values in the distribution function of the \HI effective optical depth, in agreement with recent observations. We find tentative evidence that considering some QSO emission at redshift $z \gtrsim 5$ could reconcile such observations with numerical simulations that have so far failed to reproduce
this feature. 
Finally, we show that collecting more \HeII-absorption spectra at $z\gtrsim 3$ and studying their distinctive characteristics will be key to precisely constraining the QSO contribution to reionization.
\end{abstract} 

\begin{keywords}
radiative transfer -- intergalactic medium -- cosmology:theory -- large-scale structure of the Universe -- quasars:general
\end{keywords}

\section{Introduction}
Over the last decades, a standard picture has emerged for the epoch
of cosmic reionization (EoR, see e.g. \citenp{Zaroubi2012rev}; \citenp{McQuinn2015rev} for a review).
In this scenario,
ultraviolet (UV) radiation produced by star formation in faint galaxies is responsible for the ionization of hydrogen and for the first ionization of helium in the intergalactic medium (IGM) at redshift $6 \lesssim z \lesssim 10$. Later on, at $2 \lesssim z \lesssim 4$,
harder radiation from quasars (QSOs) causes the second ionization of the diffuse helium.

This standard picture is supported by observations and theoretical considerations.
The rapid evolution in the transmission of the
Lyman-$\alpha$ (\Lya) forest at
$z \lesssim 6$ 
\citep{Fan+2006,Becker+2013,Becker+2015}, and the drop in the number density of \Lya emitters and \Lya-bright galaxies 
at $6 < z < 7$ (\citenp{Ota+2010}, \citenp{Pentericci+2011}, \citenp{Shibuya+2012}, \citenp{Furusawa+2016}, \citenp{Mason+2018}; but see \citenp{Sadoun+2017})
set constraints on the timing of hydrogen reionization which are also
supported by
the latest data on the Thomson optical depth
of the cosmic microwave background \citep[CMB,][]{Planck2016tau, Planck2018cosmo}.
Similarly, the \HeII \Lya forest 
encodes information about
`helium reionization', a conventional name
used to indicate
the transition from \HeII to \HeIII.
Although only a handful of
`clean' sightlines (with little
foreground absorption down to 
the \HeII \Lya resonance wavelength) 
are available, they consistently
show a rapid increase in the transmitted flux between $2.7 \lesssim z \lesssim 3$ \citep{Shull+2010,Furlanetto+2010,Worseck+2011,Worseck+2016,Worseck+2018}. 
Further constraints can be obtained from the evolution of the IGM temperature at mean density, inferred from the 
\HI \Lya
absorption features. 
Observational data are available only at $z \lesssim 5$ and show a large scatter, 
partially reflecting
the different data analysis techniques used to retrieve this information 
\citep[\eg][]{Rorai+2018, Boera+2018, Hiss+2018, Walther+2018}. 
Nevertheless, the peak at $z \approx 3$ is usually interpreted as a signature of the
completion of helium reionization.

Still, many details of the EoR are loosely constrained 
and there is space
for substantial modifications
to accommodate recent observations that
challenge the standard description.
The spectrum of the quasar ULAS J0148+0600 \citep{Becker+2015} contains a particularly long ($\sim 110\,\hMpc$) Gunn-Peterson trough at redshift $z=5.98$, which appears at odds \enrico{with a standard reionization history \citep[\eg][]{Chardin+2016} and may suggest an incomplete hydrogen reionization \citep{Kulkarni+2018_2}.}
More recently, \citet{Barnett+2017} observed a very extended dark gap
covering the range $6.12 < z < 7.04$ 
and corresponding to a
comoving length of
$240\,\hMpc$ with a mean \HI fraction $>10^{-4}$.
Additionally, current models of reionization
have difficulties to explain the IGM inhomogeneity indicated by the broad 
probability distribution
of the \HI optical depth
observed at $5 \lesssim z \lesssim 6$ (\citenp{Becker+2015}; \citenp{Bosman+2018}; \citenp{Eilers+2018}; but see \citenp{Gnedin+2016}). 
Several studies have
addressed this problem and indicated possible
solutions. 
It has emerged that opacity fluctuations
can be enhanced either in the presence of local temperature variations \citep{DAloisio+2015}
or by considering that 
the mean free path ($\mfp$) of ionizing photons depends on the local photoionization rate \citep{Davies+2016, Becker+2018}.
Another possibility is to
consider a scenario in which the QSO contribution to the reionization photon budget
is boosted at high redshift \citep{Chardin+2016}.
This is the direction we explore in this paper.

The rationale for our investigation
lies in an ongoing discussion in the literature regarding
the abundance of faint active galactic nuclei
at high redshift and their role during the EoR.
By applying a novel selection criterion
within a deep field with extensive  multiwavelength coverage,
\citet{Giallongo+2015} have detected an
unexpectedly large number of faint (\ie with an absolute magnitude $\MAB \sim -20$ at $1450\,\AA$) QSO candidates at $z>4$ 
\citep[but see][]{Ricci+2017, McGreer+2017, Parsa+2018, Kulkarni+2018_2}.
If confirmed, this result would 
suggest that QSOs provide an important
contribution to the photon budget during
the entire EoR and, possibly, even dominate it. Inspired by these findings, 
\citet[][\MH hereafter]{MH2015} have built an analytical model for the EoR
in which all ionizing photons are generated
by QSOs.
Interestingly, the results of the model 
satisfy the observational constraints on the evolution of the \HII fraction and on
the Thomson optical depth of CMB photons. 
Following
\citet{Giallongo+2015} and \MH, a number of 
authors have revisited the question
of the importance of non-stellar sources
during hydrogen reionization
\citep[\eg][]{DAloisio+2016, UptonSanderbeck+2016, Chardin+2016, Onoue+2017, Kulkarni+2017, Qin+2017, Mitra+2018, Hassan+2018}. However, 
these analytic or semi-numerical studies do not include
a detailed treatment of radiation transfer which is necessary
to make more accurate predictions (especially for the temperature of the IGM)
and produce realistic synthetic observations to be compared with actual data. 
On the other hand, the investigations based on fully-coupled radiation-hydrodynamical simulations consider rather small computational volumes and thus suffer from 
sample variance.
In this work, we 
improve upon existing results
by performing detailed hydrodynamical simulations of a scenario in which cosmic reionization is driven only by quasars. 
We use a suite of large simulation boxes post-processed with a radiative-transfer (RT) code  in order to track the detailed evolution of the IGM. We then produce realistic synthetic observations and use them
to (i) test the 
plausibility of the QSO-only reionization model, (ii) characterise
the impact of the QSO contribution on the Lyman-$\alpha$ forest, 
and (iii) provide predictions for a number of observables that should be able to discriminate between the standard reionization scenario and a QSO-dominated one. 
Such information is extremely valuable in order to disentangle the role of different types of sources and shed light on the properties of the high-redshift IGM. 

The paper is organized as follows.
In Section~\ref{sec:methods}, we describe
our numerical techniques and the specifics of the runs.
The simulation outputs are presented in Section~\ref{sec:results} and analysed in Section~\ref{sec:spectra} where we discuss
several mock observations that we compare
with actual data and previous theoretical work.
Finally, we summarise our findings and draw conclusions in Section~\ref{sec:conclusions}.

\section{Numerical Methods}
\label{sec:methods}

In this Section, we describe the setup of our numerical simulations, together with the modelling of the radiation sources and their calibration against recent observations. 
The techniques we use here are 
based on \citet[][hereafter \CCPone]{Compostella+2013} and \citet[][\CCPtwo]{Compostella+2014},
to which we refer for further details.

\subsection{Hydrodynamical simulations}

We consider a flat $\Lambda$CDM cosmological model 
and use the results of the \Planck
satellite to fix the parameters that determine
its background evolution and the power spectrum of the Gaussian linear perturbations \citep{Planck2015cosmo}.
For the present-day values of the
matter density, the baryon density and the Hubble parameter
we thus use $\Omega_{\rm m} = 0.306$, $\Omega_{\rm b} = 0.0483$ and $H_0 = 67.9 \, {\rm km \, s^{-1} \, Mpc^{-1}}$, respectively. 
Moreover,
the normalization of the linear power spectrum and the primordial spectral index are $\sigma_8 = 0.815$ and $n = 0.958$.

We run four hydrodynamical simulations 
using the adaptive mesh refinement (AMR) code \ramses \citep{Ramses} and employing a cubic box of comoving side $L_{\rm box} = 100 \, \hMpc$ with periodic boundary conditions. The dark matter (DM) is sampled using $256^3$ particles (corresponding to a particle mass $m_{\rm DM} = 4.3 \times 10^9\, \hMsol $) while the hydrodynamical equations are solved on a base grid of $256^3$ elements with up to 7 levels of refinement. 
This way, the simulations
reach a maximum nominal resolution of approximately $3 \, \hkpc$ and resolve the Jeans length of the gas with several computational mesh cells.
The refinement strategy is quasi-Lagrangian, namely a cell is split whenever the enclosed mass exceeds the critical threshold of $8 \, m_{\rm DM}$. 
The initial conditions 
are produced using the GRAFIC package \citep{GRAFIC}. 
The gas is assumed to follow an ideal equation of state with adiabatic index $\gamma = 5/3$ and has a primordial composition (\ie the helium mass fraction is $Y = 0.24$).
We do not track
star formation and neglect stellar feedback since the scales of interest for the
analysis of the EoR are much larger than those affected by such phenomena (see also \CCPone, \CCPtwo). 
This approximation may somewhat enhance the occurrence of dense systems that act as sinks of radiation.
Note that our simulations do not include a background of UV radiation. The reason is twofold. First, we do not consider ionizing
photons emitted by stars.
Second, the ionizing radiation produced by QSOs is
treated in post-processing as described in Section~\ref{sec:rt}.

DM haloes are identified using the \texttt{HOP} finder \citep{HOP} 
in its default configuration. We only consider haloes containing more than 70 DM particles (corresponding to halo masses 
$\Mh > 3 \times 10^{11} \hMsol$)
whose abundance agrees well with popular
fitting functions \citep[e.g.][]{JenkinsHMF}.

\subsection{Radiative transfer}
\label{sec:rt}
Modelling the timing and the properties of the EoR
requires an accurate treatment of RT. 
This consists of two parts: describing the properties of the sources of UV radiation 
and propagating the ionizing photons through the IGM.

\subsubsection{Ionizing sources}
\label{sec:sources}

\begin{figure}
\begin{center}
\includegraphics[width=\columnwidth]{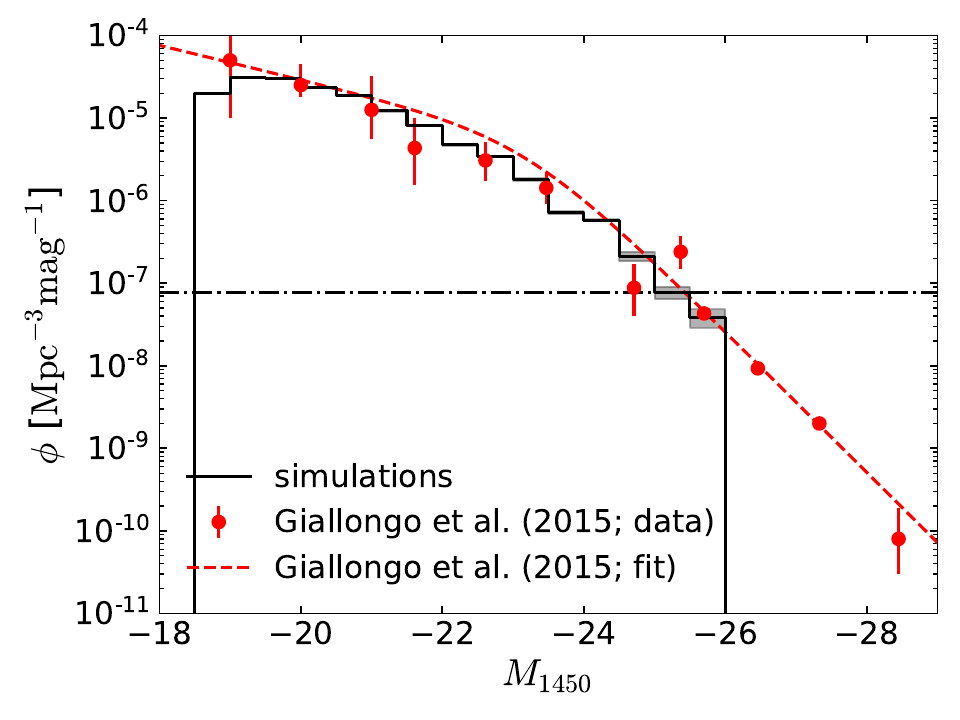}
\end{center}
\caption{The QLF, $\phi$, constructed from our simulations (solid histogram) is compared with the observational results by \protect\citet[][points with errorbars]{Giallongo+2015} and the best-fit double power law reported in the same work (dashed line). 
The shaded regions show the rms scatter of the four simulations around the mean.
The horizontal dot-dashed line highlights the value of $\phi$ corresponding to one object per simulation box and magnitude bin.}
\label{fig:LF}
\end{figure}

The spatial distribution of the ionizing sources within the cosmic web influences the development of cosmic reionization.
Having the possibility to model this effect in
a realistic way represents one of the main advantages of numerical simulations with respect to analytical models that describe radiation as a uniform background.
In order to assign a position to each source
of radiation, we assume that
QSOs reside at the centre of DM haloes 
(neglecting multiple occupancies)
and that their luminosity scales (statistically) with the mass of the host haloes
\citep[\eg][]{Silk+1998, Wyithe+2002, Volonteri+2006, Compostella+2013, Compostella+2014}. 
Several observations support this hypothesis \citep[\eg][]{Kim+2009, Cantalupo+2012, Morselli+2014, Ota+2018, Onoue+2018} although 
some exceptions have been found
\citep{Banados+2013,Fanidakis+2013,Mazzucchelli+2017}. 
In practice, we consider each DM halo
as a potential host of a QSO and sample
the associated
magnitude at $1450 \, \AA$ ($\MAB$) from a Gaussian distribution with mean
\begin{equation}
\label{eq:MAB}
\MAB = - \frac{10}{3} \log \left( \frac{\Mh}{\hMsol} \right) + \varepsilon
\end{equation}
and standard deviation $\sAB$. 
We determine the free parameters $\varepsilon$ and $\sAB$ by fitting
the shape of the quasar luminosity function
(QLF) measured by \citet{Giallongo+2015} at redshift $4.0<z<4.5$. Using 1000 realisations of the QSO assignment, we obtain $\varepsilon = 19$ and $\sAB = 1.25$ (with some freedom within a degeneracy region in parameter space).
This implies that the resolved haloes
in our simulations host QSOs in the range
$-25.76\leq \MAB \leq -18.67$. Our ionizing sources thus sample the faint end of the QLF which generates the majority of the ionising photons (only 11.6 per cent of the total should be emitted by the unresolved faint sources). 

We employ a lightbulb model to describe
the QSO activity. 
Sources are randomly switched on with a probability that is independent of the properties of the host halo.
Once activated, a QSO has a constant
emissivity for a lifetime of 45 Myr and is switched off afterwards. Each
source can become active multiple times during a simulation run.
The resulting duty cycle is consistent
with many observational studies which, 
however, set only weak constraints
\citep{Jakobsen+2003, Martini2004, Porciani+2004, Cantalupo+2007, Worseck+2007, Kelly+2010, Bolton+2012, Cantalupo+2012, Trainor+2013, Borisova+2016, Schmidt+2017}. 

The fraction of active sources at $z\sim 4$ is determined by matching the amplitude of the QLF measured by \citet[][see Fig. \ref{fig:LF}]{Giallongo+2015}.
Its redshift dependence, instead, is determined by adopting a `pure density evolution' model, as follows.
First, we assume that the shape of the QLF does not evolve (which is
consistent, to first approximation, with the results in \citenp{Giallongo+2015}).
Second, we change the fraction of active sources so that to match the evolution of the emissivity
at $912 \, \AA$, $\epsilon_{\rm 912}(z)$, obtained by \MH after extrapolating down to $z=12$ a collection of observational data at $z\lesssim 6$.
In practice, this is done by
changing the number of active sources, $N_{\rm src, active}(L,z)$, according to the relation
\begin{equation}
\label{eq:Nsrc_resc}
N_{\rm src, active} (L,z) = C(z) \, N_{\rm src,active} (L,z=4)\;,
\end{equation}
where
\begin{equation}
\label{eq:Nsrc_norm}
C(z) = \epsilon_{\rm 912}(z) / \epsilon_{\rm 912}(z=4)\;.
\end{equation}
This procedure ensures that our
integrated emissivity exactly matches
the input used in the analytical model by \MH. On average, we end up having between 120 and 400 active sources within a single simulation box, the number increasing with time.

We model the rest-frame QSO spectral energy distribution 
with a broken power law:
\begin{equation}
f(\nu) \propto 
			\begin{cases}
			\nu^{\afuv}, \, \lambda \geq \lambda_{\rm b} \\
			\nu^{\aeuv}, \, \lambda   <  \lambda_{\rm b}
			\end{cases}
\end{equation}
where $f(\nu)$ is the energy flux, $\afuv$ and $\aeuv$ are the spectral indices in the far UV and extreme UV, respectively, and $\lambda_{\rm b}$ is the junction point of the two power-laws. Following \citet{Lusso+2015}, we assume $\lambda_{\rm b} = 912 \, \mathrm{\AA}$ and assign to each source a pair of spectral indices sampled from two Gaussian distributions with means and standard deviations corresponding to $\afuv = -0.61 \pm 0.01$ and $\aeuv = -1.7 \pm 0.61$, respectively. 
This matches the observed spectrum of low-redshift QSOs corrected for IGM absorption.

\subsubsection{Propagation of radiation}
In order to model the radiation transport, we employ an upgraded version of the three-dimensional RT code \texttt{RADAMESH} \citep{RADAMESH}.
This software implements
an efficient photon-conserving ray-tracing algorithm and is designed for AMR grids.
\texttt{RADAMESH} uses a cell-by-cell Monte Carlo scheme to sample the radiation field at each location. 
The temperature and ionization state of each resolution element are computed using a non-equilibrium
fully-implicit chemistry solver that includes six different species (\HI, \HII, \HeI, \HeII, \HeIII and $e^{-}$).

In order to limit the computational time to an acceptable amount, we artificially decouple
the RT from the hydrodynamic evolution of the gas.
We thus post-process the $z=4$
snapshot of the hydro simulations
with the RT code.
Ionizing photons emitted from the discrete sources described in Section~\ref{sec:sources}
are propagated through the simulation box.
The gas density, temperature and ionization states are updated 
keeping into account the local
photoionization, photoheating,  recombination and cooling rates as well as the expansion of the Universe.
However, the spatial pattern of the density fluctuations is fully determined by the $z=4$ output of the hydro simulation.
We briefly comment on the robustness of this approximate method in the next Section.

To ensure a proper comparison with the analytical study by \MH,
we assume that
the first QSOs light up at $z_{\rm start}=12$
which is compatible with the predictions of large cosmological simulations \citep[\eg][]{DiMatteo+2016}.
The RT runs are then evolved until redshift $z_{\rm end} = 3.5$, when both hydrogen and helium are completely ionized (\ie their neutral volume fraction is less than $10^{-5}$). 
 
We sample the radiation spectrum 
between 1 and 40 Ry
using 50 bins 
logarithmically spaced within three sub intervals starting at the
ionizing thresholds of \HI, \HeI, and \HeII.
In details, we use 
10 bins between $1$ and $1.81$ Ry, 10 bins between $1.81$ and $4$ Ry, and 30 bins in the range $(4,40]$ Ry. 
Further increasing the number of bins or modifying their frequency range produces only negligible changes in the gas temperature
(see Appendix B in \CCPone for a convergence test).
Note that our simulations include
soft X-rays that, thanks to their long mean free path, pre-heat the gas ahead of the ionization fronts. 
On the other hand, we neglect secondary ionizations that are expected to have only a minor impact 
\citep{Shull+1985, McQuinn+2009, Kakiichi+2016concerted}.

We have checked that extremely bright sources that are not represented in our simulation box should not introduce significant fluctuations in the distribution of the ionizing radiation. In fact, the expected number of these QSOs with a Str\"omgren sphere \citep[e.g.][]{Haiman+2001} that overlaps with the box 
is always much smaller than one
for any realistic combination of the QSO lifetime and the ionization state of the IGM at $z \gtrsim 5$.

At late times, when the gas in the simulations is almost completely ionized, the mean free path of the most energetic photons can exceed the box size. In order to reduce the computational time, we discard the photon packets that freely traverse the box two times (at variance with \CCPtwo where they were replaced by a uniform background). 
This approximation might slightly shift our results towards higher optical depths at the very end of the EoR.

\subsubsection{Post-process RT: motivations and accuracy}
RT is a very computationally intensive problem.
The intensity of radiation depends on seven variables 
(three spatial and two angular coordinates as well as time and photon energy) and the RT equation is non local. In consequence, for the current technology,
coupled RT and hydrodynamic equations
turn out to be too challenging to solve within large spatial domains without making simplifying assumptions.
Different approaches have been
followed in the literature to study the EoR.
Some authors prefer to study
the coupled evolution
at the price of introducing
crude approximations in the RT modelling (\eg monochromatic radiation) and/or considering limited volumes \citep[\eg][]{Croc,CoDa,EMMA,Aurora}.
Others employ more detailed RT algorithms  at the cost of decoupling them from the hydrodynamics \citep[\eg][]{Paschos+2007, McQuinn+2009, Rahmati+2013, Compostella+2013, Bauer+2015, Kakiichi+2016bubble}. 

In this work, we follow the second approach  
in order to simulate
a representative cosmological volume
and suppress random fluctuations
in the number of rare sources like QSOs. This strategy also
offers us the possibility of comparing 
our results with the analysis of the standard EoR scenario presented in 
\CCPone and \CCPtwo using the same
numerical setup.

It has been shown that 
decoupling the RT from the hydrodynamical response of the gas
has a small impact on the models for the EoR.
For instance, \citet{Meiksin+2012} found
that gas velocities are typically
altered by $\sim 1\, \kms$ 
while gas densities change by
less than 10 per cent. 
Similar conclusions have been reached by employing the same codes and setup used in this work (see Appendix A in \CCPone): the gas 
density is altered by less than
$\sim 5$ per cent down to $z=3.2$ with the largest deviations seen
in filamentary regions around
mean density. We are thus confident that our result are sufficiently robust and accurate.

\begin{figure*}
\begin{center}
\includegraphics[width=\textwidth]{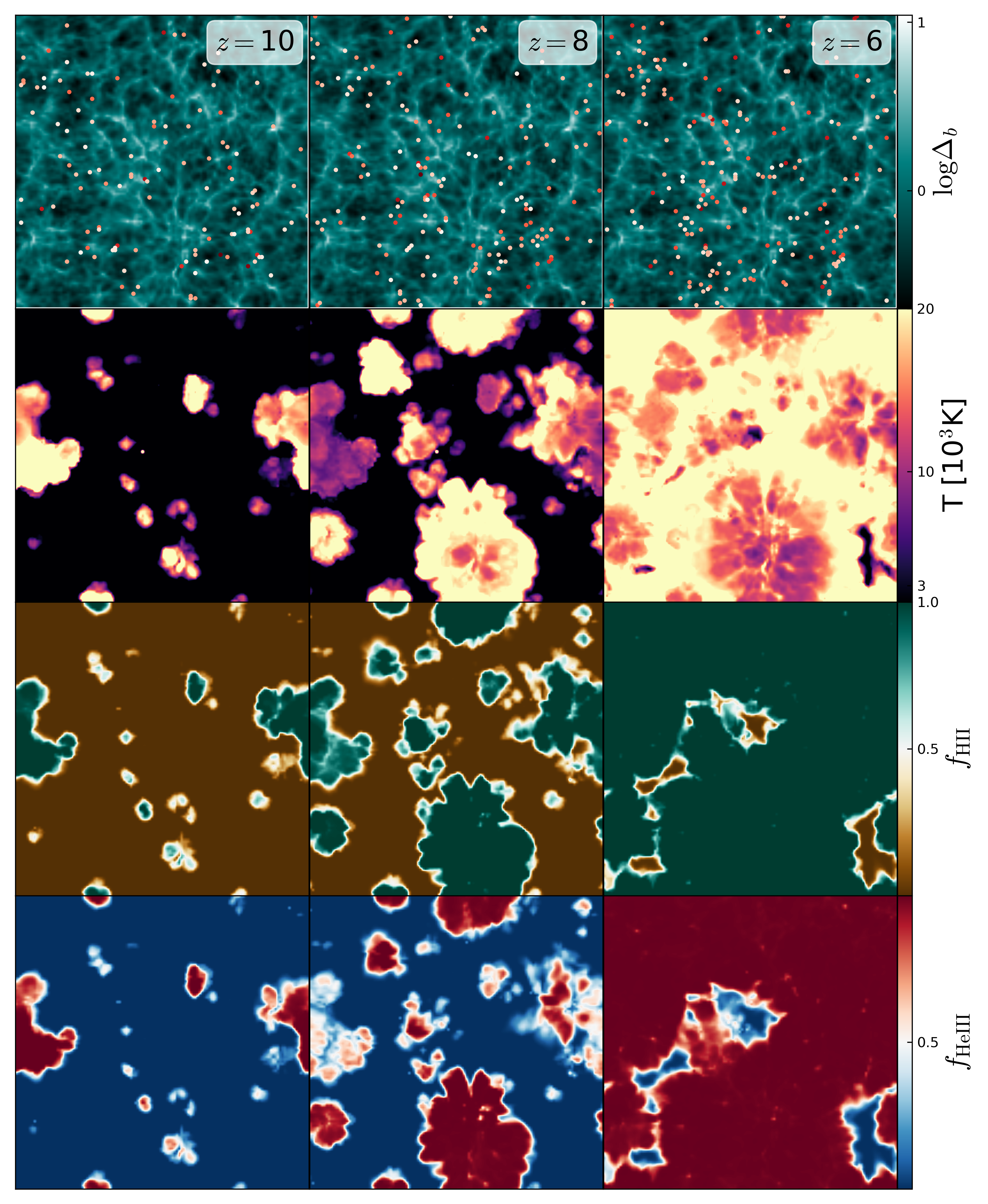}
\end{center}
\caption{
Simulation snapshots referring to a thin slice passing through the centre of one of our boxes.
The time evolution proceeds from left to right across the frames
that correspond to $z=10, 8$ and 6.
In the top row, the baryon overdensity is shown together with
the projected position of the active sources
located within $\sim 4\, \hMpc$ from the slice (redder colors denote brighter sources). 
The second, third and fourth rows display the gas temperature
as well as the \HII and \HeIII volume fractions, respectively.}
\label{fig:mosaic}
\end{figure*}

\begin{figure*}
\begin{center}
\includegraphics[width=\textwidth]{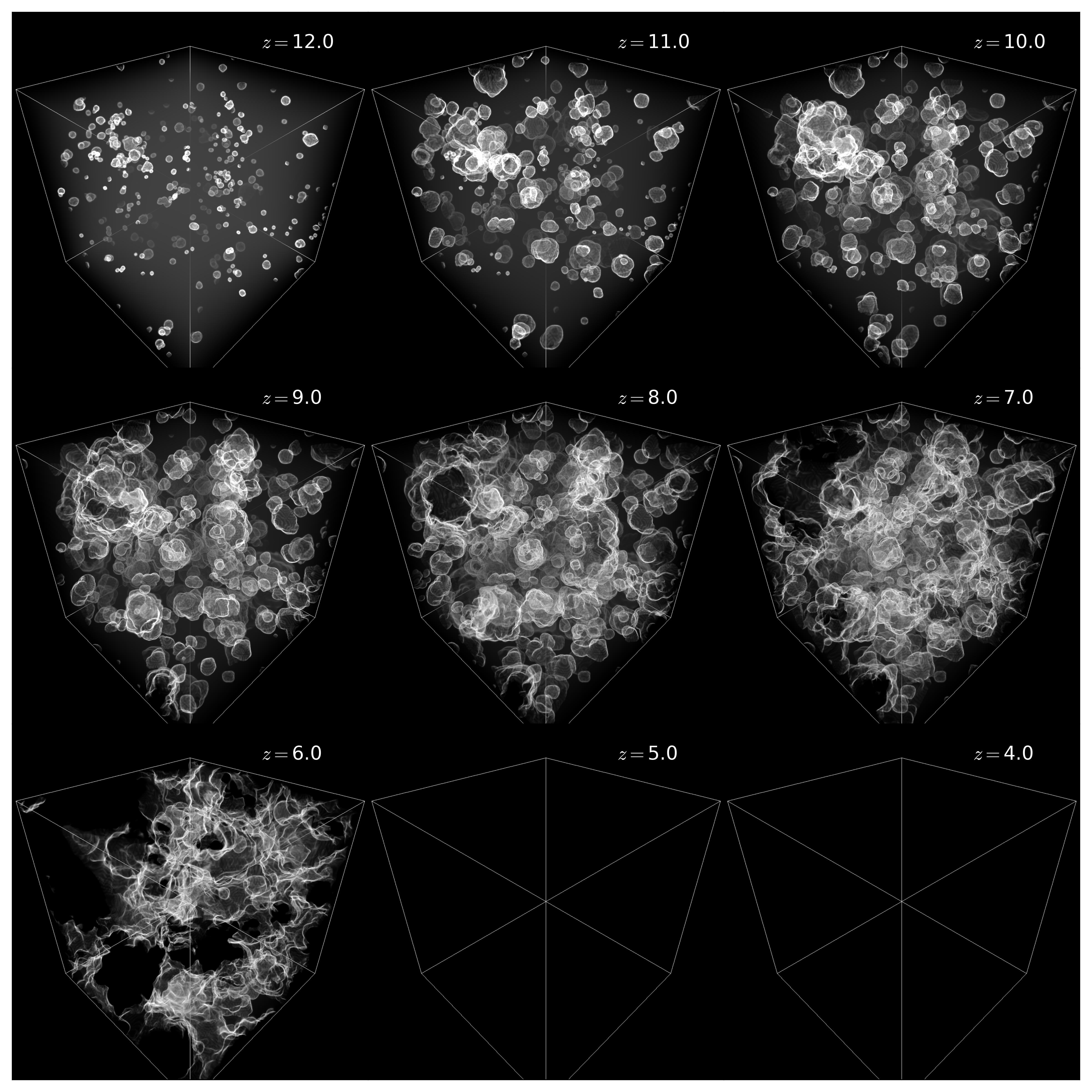}
\end{center}
\caption{Volume rendering of the \HI ionization front in one of our simulations at different redshifts. Neutral gas is shown in semi-transparent white, while the ionization fronts (defined as the regions where the local \HII fraction is 50 per cent) 
are shown using opaque white.}
\label{fig:collage}
\end{figure*}

\section{Results}
\label{sec:results}

In order to give a visual impression of the development of the EoR, 
Fig.~\ref{fig:mosaic} shows
a series of snapshots extracted from one of our simulations and displaying
the redshift evolution of different
physical quantities.
From left to right, columns refer
to $z=10, \, 8,$ and $6$, respectively.
The top panels display the baryonic
overdensity $\Delta_{\rm b} (\mathrm{x})\equiv \rho_{\rm b} (\mathrm{x}) / \bar{\rho}_{\rm b}$, where $\rho_{\rm b} (\mathrm{x})$ is the baryon density and $\bar{\rho}_{\rm b}$ denotes its mean value within the box. 
Over plot are the projected positions of the nearby active sources, color-coded in such a way that redder colors correspond to brighter QSOs.  
The second row of panels illustrates changes in the gas temperature, while the last two exhibit the evolution of the \HII and \HeIII fractions, respectively.
Note that, contrary to what happens in the standard scenario, hydrogen and helium get fully ionized simultaneously at a given location.  When dominated by QSOs,
cosmic reionization 
proceeds in a very inhomogeneous fashion. First, individual ionized
bubbles are formed that then percolate.   

This aspect is further elucidated
in Fig.~\ref{fig:collage} where we show a time sequence
of volume renderings of 
the \HI ionization fronts (here defined as the regions where the local \HII fraction is 50 per cent).
A topological change due to the percolation transition is clearly noticeable at $z\sim 6$.

\subsection{Ionized fractions}
\label{sec:ion_frac}

\begin{figure}
\begin{center}
\includegraphics[width=\columnwidth]{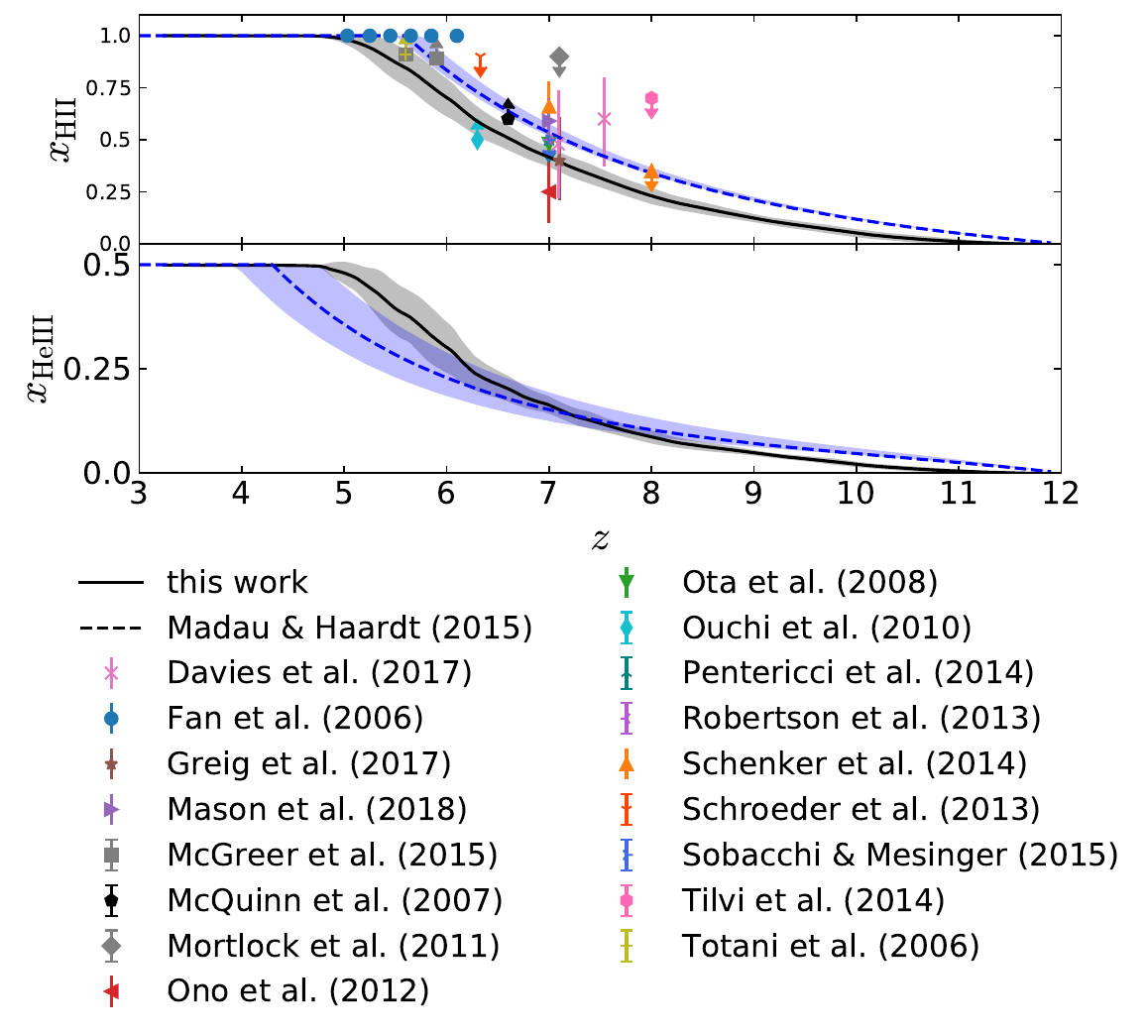}
\end{center}
\caption{Volume fraction of \HII (top) and \HeIII (bottom) computed from our simulation suite. The solid lines and the surrounding shaded regions show the mean and the rms scatter among the four realizations. The dashed lines indicate the predictions by \MH and the shaded areas around them describe the effect of varying the parameters of their analytical model. 
The symbols correspond to a collection of constraints on the hydrogen ionized fraction
as indicated by the labels.}
\label{fig:ionised_fraction}
\end{figure}

In Fig.~\ref{fig:ionised_fraction}, we provide a first quantitative evaluation of
the impact of early QSOs on the EoR by
studying the redshift evolution of
the ionized volume fraction (often referred to also as the `volume filling factor')
for \HII ($x_{\HII}$, top panel) and \HeIII ($x_{\HeIII}$, bottom panel). We do not show the \HeII fraction here (nor we discuss it elsewhere in the paper) as it matches almost perfectly $x_{\rm \HII}$ as a consequence of the hard radiation spectrum emitted by QSOs and
the close first-ionization energies of 
hydrogen and helium.
The solid line indicates the average over our simulations 
and the surrounding light shaded region denotes the corresponding rms value. The four runs are in very good agreement and the scatter among them is small. 
Different symbols indicate a number of observational
constraints on the \HII fraction \citep{Fan+2006, Totani+2006, McQuinn+2007, Ota+2008, Ouchi+2010, Mortlock+2011, Ono+2012, Schroeder+2013, Robertson+2013, Schenker+2014, Pentericci+2014, Tilvi+2014, McGreer+2015, Sobacchi+Mesinger2015, Greig+2017, Mason+2018,Davies+2018}.
This confirms that the QSO-dominated model produces enough photons to generate an EoR
and keep the IGM ionized afterwards
as suggested by previous analytical work.
An obvious benchmark for our simulations is the \MH \ model 
(dashed line) which predicts similar
volume fractions.
Some important differences can nevertheless be appreciated: in our simulations, reionization proceeds slower at $z \gtrsim 9$ and becomes faster at $z \lesssim 7.5$. The reason is that \MH
describe the ionizing photons 
as an uniform UV background
and do not consider the precise location of the sources as well as RT effects.
However, our simulated QSOs
reside in highly overdense regions of the Universe that are characterized by a faster
recombination rate than average. The net effect is to slow down the progression of the reionization process around the active sources.
Later on, when the overdense patches are completely ionized, the ionization fronts reach underdense regions, in which 
reionization 
takes place faster than in the analytical model. 

\begin{figure}
\begin{center}
\includegraphics[width=\columnwidth]{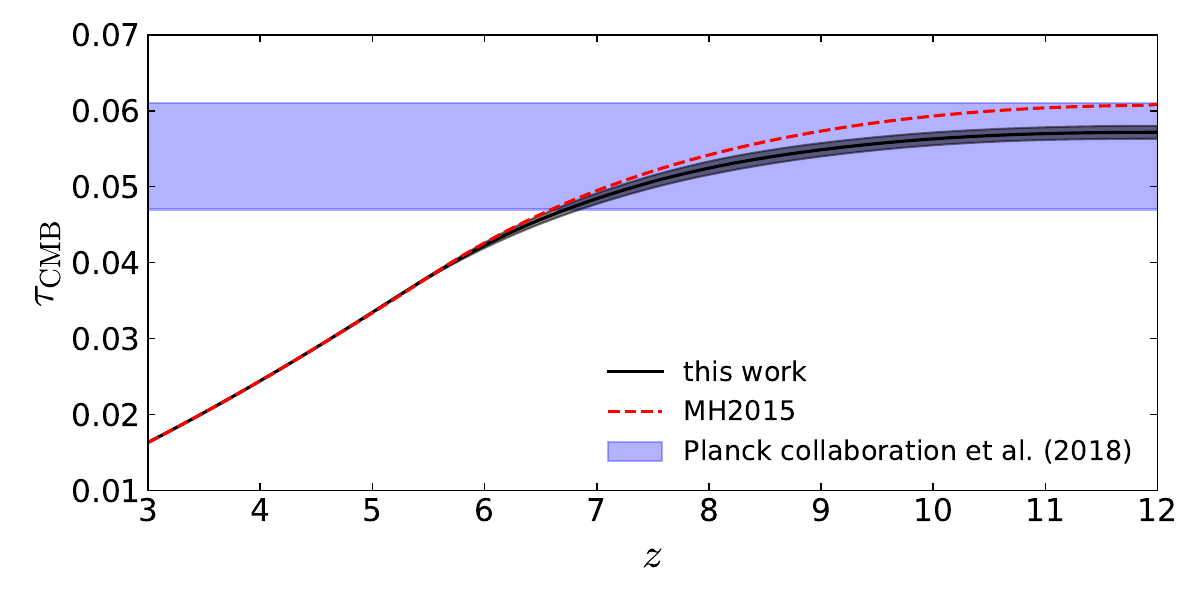}
\end{center}
\caption{Optical depth to Thomson scattering integrated from the present
time to redshift $z$.
The solid line and the surrounding shaded region correspond 
to the results from our simulations shown in Fig.~\ref{fig:ionised_fraction}.
The dashed curve shows the prediction by \MH. The large shaded area highlights the current observational constraints for the CMB at $z\sim 1100$ \citep{Planck2018cosmo}.}
\label{fig:tau_CMB}
\end{figure}

As we have briefly mentioned above,
a striking 
feature characterizing the QSO-dominated
scenario (compared with
the standard model of the EoR) 
is the nearly simultaneous ionization of \HI and \HeII. 
This 
is a direct consequence of the assumption that only one population of sources provides all the ionizing photons for both species. 
By directly comparing the evolution of $x_{\rm \HII}$ and $x_{\rm \HeIII}$ in Fig.~\ref{fig:ionised_fraction}, it emerges
that the volume fraction of ionized hydrogen is a bit higher at all times. This
small delay in the reionization of helium reflects
the shape of the QSO spectrum 
that
gives less ionizing photons per helium atom than for hydrogen 
(see, however, the discussion in Section \ref{sec:conclusions} on the impact of the escape fraction).

The peculiar reionization histories in the QSO-only scenario modify the number
density of free electrons in the IGM with
respect to the standard model of the EoR.
In Fig.~\ref{fig:tau_CMB}, 
we show that 
the resulting optical depth of CMB radiation to Thomson scattering, $\tau_{\rm CMB}$,
still lies within the observational constraints \citep{Planck2018cosmo}
as also derived by \MH (dashed lines).

\subsection{IGM temperature}
\label{sec:temp}

\begin{figure}
\begin{center}
\includegraphics[width=\columnwidth]{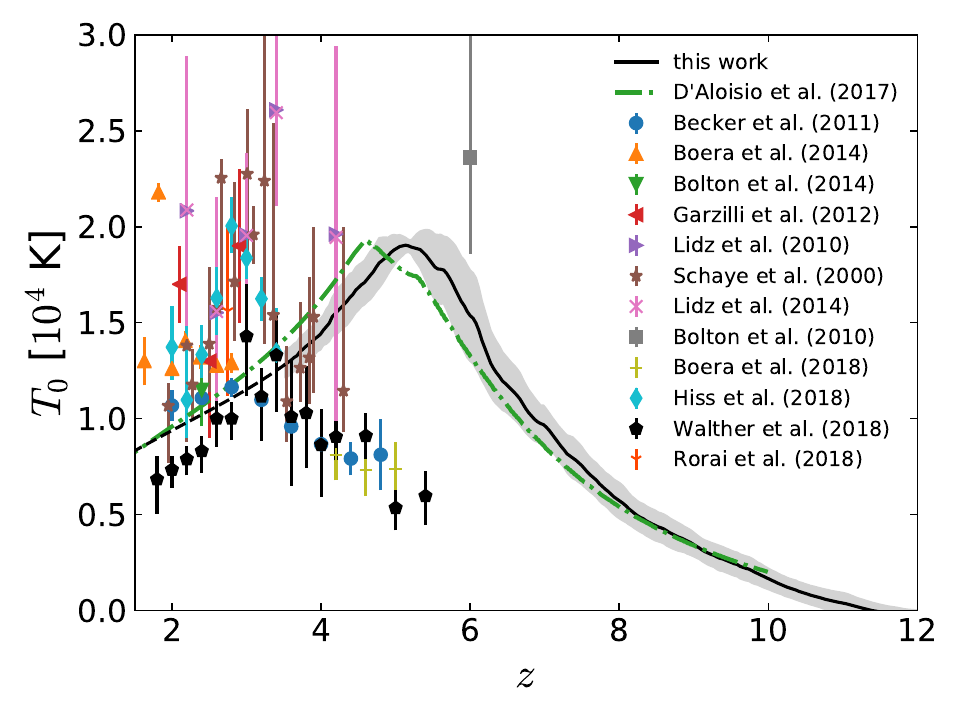}
\end{center}
\caption{Volume-weighted average temperature of the IGM at mean density, $\Tzero$, as a function of redshift. The average (solid line) and the rms scatter (shaded area) are evaluated
over the entire simulation suite.
For $z<3.5$, we use an analytical
approximation to extrapolate the evolution
of $\Tzero$ beyond the range covered by the simulations (dashed line). 
The symbols show a collection of observational data (see the main text for details).
The dot-dashed line represents the semi-analytical prediction by \citet{DAloisio+2017}.}
\label{fig:T0}
\end{figure}

Although analytic models of reionization
can compute the ionized volume fractions
rather accurately, they cannot make robust predictions for the IGM temperature which is heavily influenced by RT effects.
For instance, \MH assume that the IGM has a uniform temperature of $T_{\rm IGM} = 2 \times 10^4$ K, neglecting any dependence on redshift or density. 
In Fig.~\ref{fig:T0},
we show the redshift evolution 
of the 
gas temperature at mean density ($\Tzero$) in our simulations.
The solid line and the shaded area around it denote the mean and the associated scatter among all the mean-density cells in the four realizations.
For completeness, we extrapolate $\Tzero$ at $z<3.5$ (dashed line) 
as in \citet{McQuinn+2016} 
by taking into account that, when the
gas is fully ionized, the temperature at mean density is determined by the adiabatic expansion of the Universe (plus smaller contributions due to Compton and free-free cooling). 
Overall, the IGM is photo-heated 
until $z\simeq 5.5$ and cools down afterwards.
Note that, excluding its maximum value,
$\Tzero\ll 2 \times 10^4$ K at all times.
A similar trend was derived by \citet{DAloisio+2017} using semi-analytical methods (although, in their case, the peak temperature is reached a bit later, at $z\sim 4.5$, see the dot-dashed line in Fig.~\ref{fig:T0}).

Several observational constraints 
published by various authors using substantially different methods 
are overplotted in Fig.~\ref{fig:T0}.
Namely, we show the results obtained by
\citet[][using the distribution of line widths in the \Lya forest]{Schaye+2000}, \citet[][from the Doppler parameter in the quasar proximity zone]{Bolton+2010}, \citet[][from the curvature statistic]{Becker+2011}, \citet[][via a wavelet filtering analysis]{Garzilli+2012}, \citet[][from the distribution of line widths]{Bolton+2014}, \citet[][employing the curvature statistic]{Boera+2014}, \citet[][using the Morlet wavelet filter analysis]{Lidz+2014}, 
\citet[][from the joint distribution of \HI column densities and Doppler parameters]{Rorai+2018}, \citet[][using the flux power spectrum]{Boera+2018}, \citet[][employing the cutoff in the Doppler parameter distribution]{Hiss+2018} and \citet[][using the \Lya forest power spectrum]{Walther+2018}.
Although statistical errorbars tend to be large and different methods do not always match, 
the available data
show a peak around $z \approx 3.5$, which is usually interpreted as a signature of late helium reionization \citep[\eg][ and references therein]{McQuinn2015rev}.
This clearly poses
a severe challenge to the QSO-dominated model
which cannot accommodate such a late
local maximum in $\Tzero(z)$.
In general, it is impossible to fit the existing temperature constraints if \HeII reionization takes place at $z>4.5$ \citep{UptonSanderbeck+2016, DAloisio+2016}.

\section{Synthetic observations}
\label{sec:spectra}

The hydrogen and helium \Lya forests are
powerful probes of the IGM properties.
In this Section, we use our simulations
of the QSO-dominated
reionization model 
to produce synthetic absorption-line
spectra for \HI and \HeII\ that we then
compare with observational data and previous numerical studies. Finally, we suggest new ways to analyse the experimental data and better 
constrain the QSO contribution to reionization.

The mock spectra are generated as in \CCPone. In brief, we compute the IGM
absorption profile as a function of wavelength by keeping into account the effects
of density, temperature and velocity.
For each simulation snapshot, 
we consider 100 random lines of sight, each one extending for $100 \, \hMpc$.
In total, we produce $\sim 12000$ spectra
with an initial resolution of $1\,\kms$ that we subsequently degrade by using a Gaussian filter with a full width at half maximum of $88\,\kms$ in order to mimic the instrumental response of an actual spectrograph. The smoothing length matches the nominal resolution of our simulations in low-density regions at $z\simeq 3.5$. Although our mock spectra do not resolve individual absorption features in the Ly$\alpha$ forest, they do encode information about the IGM opacity.

\subsection{Effective optical depths}
\label{sec:taus}

\begin{figure}
\begin{center}
\includegraphics[width=\columnwidth]{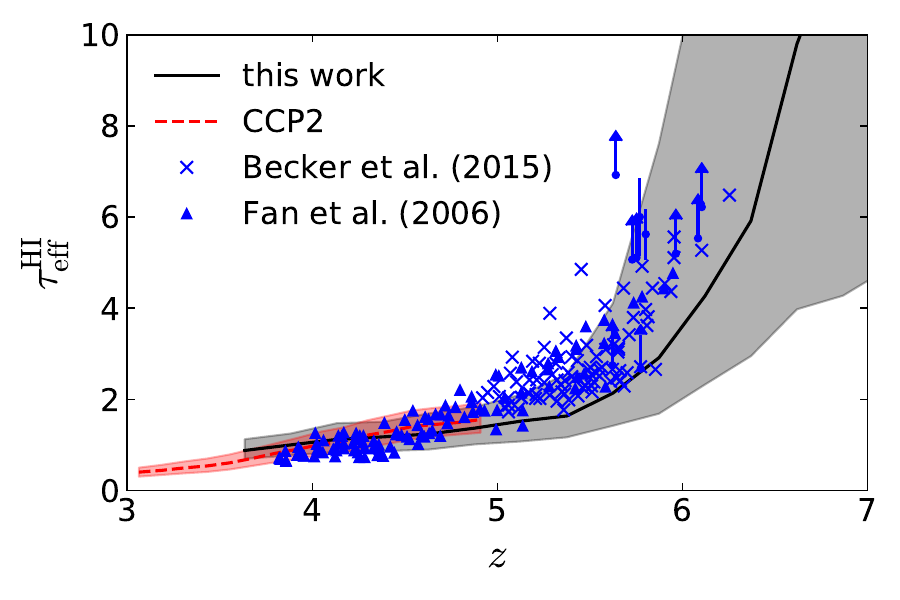}
\end{center}
\caption{The redshift evolution of the \HI effective optical depth 
computed from synthetic spectra in different numerical simulations is compared with recent
observational estimates. 
The solid line shows the median value obtained from our simulation suite of
the QSO-dominated scenario for the EoR. The corresponding evolution in the standard model of reionization (see \CCPtwo) is displayed with a dashed line.
In both cases, the shading indicates the central 68 per cent of data. The observational results by \protect\citet{Becker+2015} and \protect\citet{Fan+2006} are shown with triangles and crosses (or vertical arrows for lower limits), respectively.}
\label{fig:tau_eff_HI}
\end{figure}

The evolution of the effective optical depth, $\taueff$, has emerged as one of the most widely used characterizations of the \Lya forest.
In order to evaluate this statistic for our simulations, we first divide 
each synthetic spectrum
in chunks of size $\Delta z = 0.1$ \citep[as in \eg][]{Becker+2013}
and compute the (continuum-normalized) mean transmitted flux in it, $0\leq \langle F\rangle \leq 1$.
The effective optical depth is then obtained using  $\taueff = - \ln \langle F \rangle$.

The resulting values for the \HI forest 
are plotted in Fig. \ref{fig:tau_eff_HI}.
The solid line shows the median value in each redshift bin and the surrounding shaded region encloses the central 68 per cent of the data. 
For comparison, we also display the results obtained by \CCPtwo within the standard scenario for reionization (dashed line and shaded region). 
The large scatter seen in our simulations at $z \gtrsim 5.5$ 
is a clear indication of patchy reionization 
caused by
the low number density of QSOs 
(see also Section \ref{sec:PDF}).
Overplotted are observational data for
42 quasar spectra \citep{Fan+2006,Becker+2015}.
At $z \lesssim 4.5$, the simulations are in excellent agreement with the observations, 
especially taking into account that 
they should be
slightly biased towards higher effective optical depths
at low redshifts (see Section~\ref{sec:rt}).
On the other hand, at higher redshifts,
the IGM in the simulations tends to be too transparent, with most of the observational data falling in the upper half of the synthetic distribution of $\taueff^{\rm HI}$.

\begin{figure}
\begin{center}
\includegraphics[width=\columnwidth]{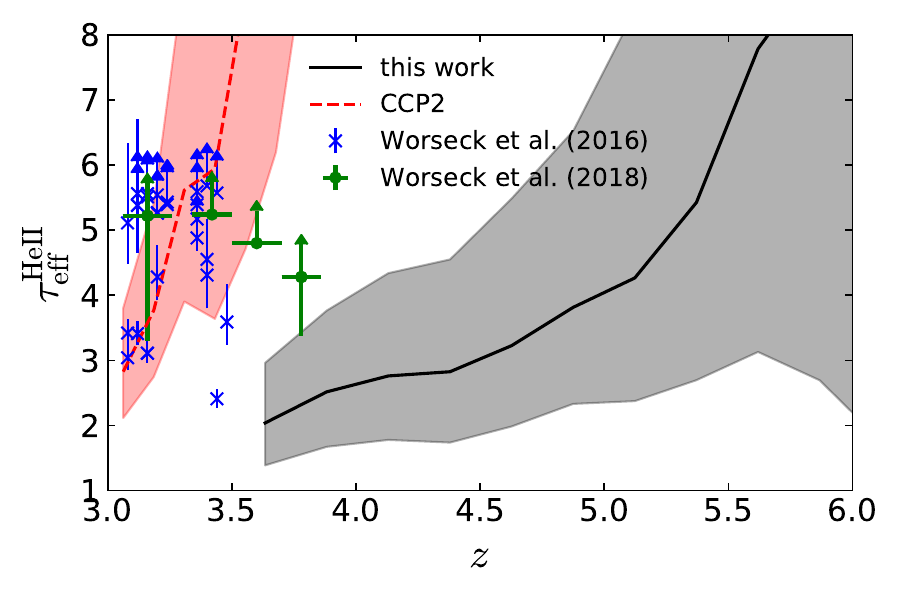}
\end{center}
\caption{As in Fig.~\ref{fig:tau_eff_HI} but for the \HeII effective optical depth. 
The data points indicate the measurements by \citet[][crosses]{Worseck+2016} and 
\citet{Worseck+2016}. For visual clarity only the median and 68 per cent central data within each redshift bin are shown]{Worseck+2018}.}
\label{fig:tau_eff_HeII}
\end{figure}	

We repeat the same analysis for
the \HeII \Lya forest
but using chunks with $\Delta z=0.04$
as in \citet{Worseck+2016}.
The late evolution\footnote{In the QSO-dominated model, $\taueff^{\rm HeII}$ changes
in a peculiar way with redshift. After the first QSOs become active, \HeII is present
only in between the hydrogen and helium ionization fronts which are generally close
in space. Therefore, the effective optical depth assumes rather low values at early times
($\taueff^{\rm HeII} \lesssim 6$)
that steadily grow as the separation between the hydrogen and helium fronts
increases due to the steep spectral index
of the ionizing radiation.
The effective optical depth reaches its maximum value around the epoch of hydrogen reionization and decreases afterwards. In fact, once the \HII bubbles percolate, \HeII regions find themselves illuminated by multiple sources and are rapidly turned into \HeIII. }
of $\taueff^{\rm HeII}$ is plotted
in Fig. \ref{fig:tau_eff_HeII}, 
together with
recent measurements from \citet{Worseck+2016} and \citet[][including new and re-analyzed spectra]{Worseck+2018}. 
Despite the small number of experimental data in the redshift range covered by our simulations,
the model and the observations are in strong tension.
In fact, the optical depths predicted by the QSO-dominated scenario at $z\approx 3.7$ are substantially lower than the observed ones at $z=3.4$. 
On the contrary, the standard reionization scenario 
is in good agreement with the available data.

In summary, the QSO-dominated scenario we have simulated
does not match the observed evolution of the IGM opacity.
For what concerns \HI, however, a moderate delay in the appearance of the first active sources and/or a steeper evolution of the ionizing emissivity could reduce (and likely completely remove) the small tension we have found with the data at intermediate redshifts.
Conversely,
it does not seem possible to reconcile the model and the data for \HeII by making small adjustments. Simply, helium reionization is completed too early in the QSO-dominated scenario.

\begin{figure*}
\begin{center}
\includegraphics[width=\textwidth]{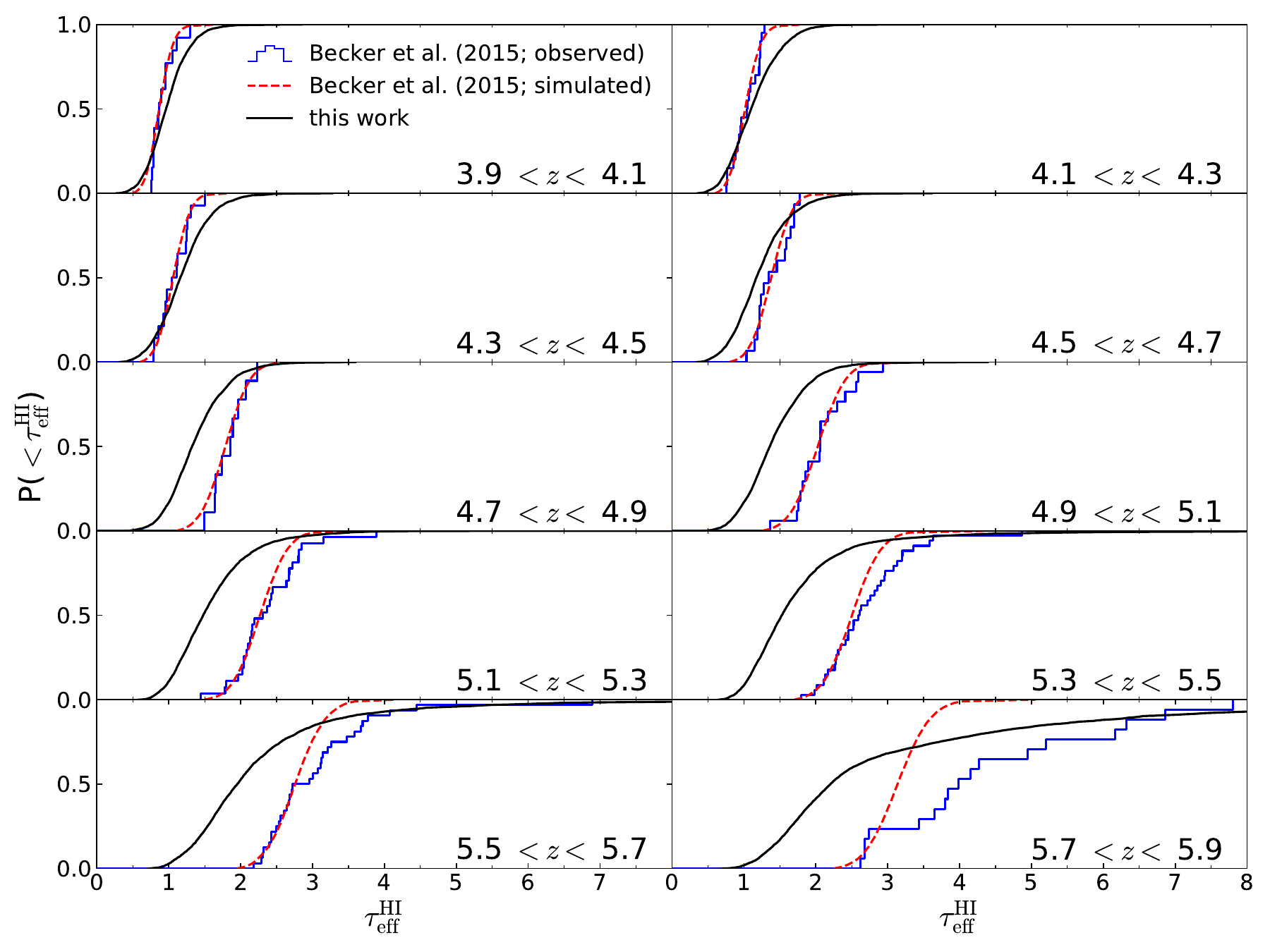}
\end{center}
\caption{Cumulative distribution function of the \HI effective optical depth. 
The solid curve shows the median over our simulations. 
The histogram displays the observational results by \protect\citet{Becker+2015}, while the dashed curve indicates the CDF predicted using numerical simulations of a standard reionization history calibrated to match the low-optical-depth end of the observed distribution (also from \protect\citenp{Becker+2015}). Each panel corresponds to a different redshift bin as indicated by the labels.}
\label{fig:CDF}
\end{figure*}

\subsection{The probability distribution function of optical depths}
\label{sec:PDF}

The spread in $\taueff^{\rm HI}$
recorded at fixed redshift indicates that
the opacity of the intergalactic hydrogen varies between different sightlines. 
The cumulative distribution function (CDF) of $\taueff^{\rm HI}$ thus provides
a simple and convenient characterization of these local changes.
In Fig.~\ref{fig:CDF}, we plot the
CDF measured by \citet{Becker+2015} 
using 
spectral chunks with a comoving length of $50\,\hMpc$.
It is well known that 
these results, that have been recently
confirmed with increased statistical significance \citep{Bosman+2018, Eilers+2018},
are inconsistent with the distribution inferred from numerical simulations
of the standard reionization model that employ a uniform UV background\footnote{Although using a smooth radiation field represents a rather crude approximation during the early phases of the EoR, it should be a sufficiently good working hypothesis after bubble percolation (\ie at $z \lesssim 6$).} 
calibrated to match the observed low-optical-depth data (see the dashed lines in Fig. \ref{fig:CDF}).
In fact, the CDF extracted from the simulations is too steep and can not reproduce the extended tail of large optical depths observed at redshift $z \gtrsim 5$
\citep{Becker+2015}.
It is thus interesting to verify whether
the QSO-dominated scenario (with its rarer ionizing sources) may help reconciling the discordance with the observations.
The solid lines
in Fig.~\ref{fig:CDF}
represent the CDF obtained from our mock spectra.
Before proceeding with the comparison, it is important to stress that
our optical depths have not been calibrated to match any part of the observed CDF.
In general, the modified scenario for the EoR does not reproduce the observations at any redshift.
At $z\lesssim 4.5$, the observed CDF is steeper than the simulated one, meaning that the IGM is more homogeneous than predicted by the QSO-dominated model. 
 
Including fainter QSOs in the simulations may relieve this tension, as a larger number density of sources can produce a more homogeneous IGM. 
At $4.5 \lesssim z \lesssim 5.5$, the simulated CDF has the correct shape but is systematically shifted towards lower optical depths (as already seen in Fig.~\ref{fig:tau_eff_HI}). At even higher redshift,
the disagreement increases at small optical depths but the synthetic CDF nicely reproduces the high-$\taueff^{\rm HI}$ tails of the distribution (mainly driven by spatial fluctuations of the photoionization rate but also by the large islands of neutral hydrogen present at these redshifts).
This is the opposite trend to that expected in the standard reionization model
\citep{Becker+2015} unless the EoR for \HI extends down to $z\sim 5.3$ \citep{Kulkarni+2018_2}.
It is therefore tempting to interpret our results as implying
that an increased QSO contribution at high redshift (with respect to the standard reionization history, see also \citenp{Chardin+2016}) might bring the theoretical predictions in agreement with observations. 
Dedicated numerical studies are necessary in order to settle this issue which is beyond the scope of this paper.

\begin{figure*}
\begin{center}
\includegraphics[width=\textwidth]{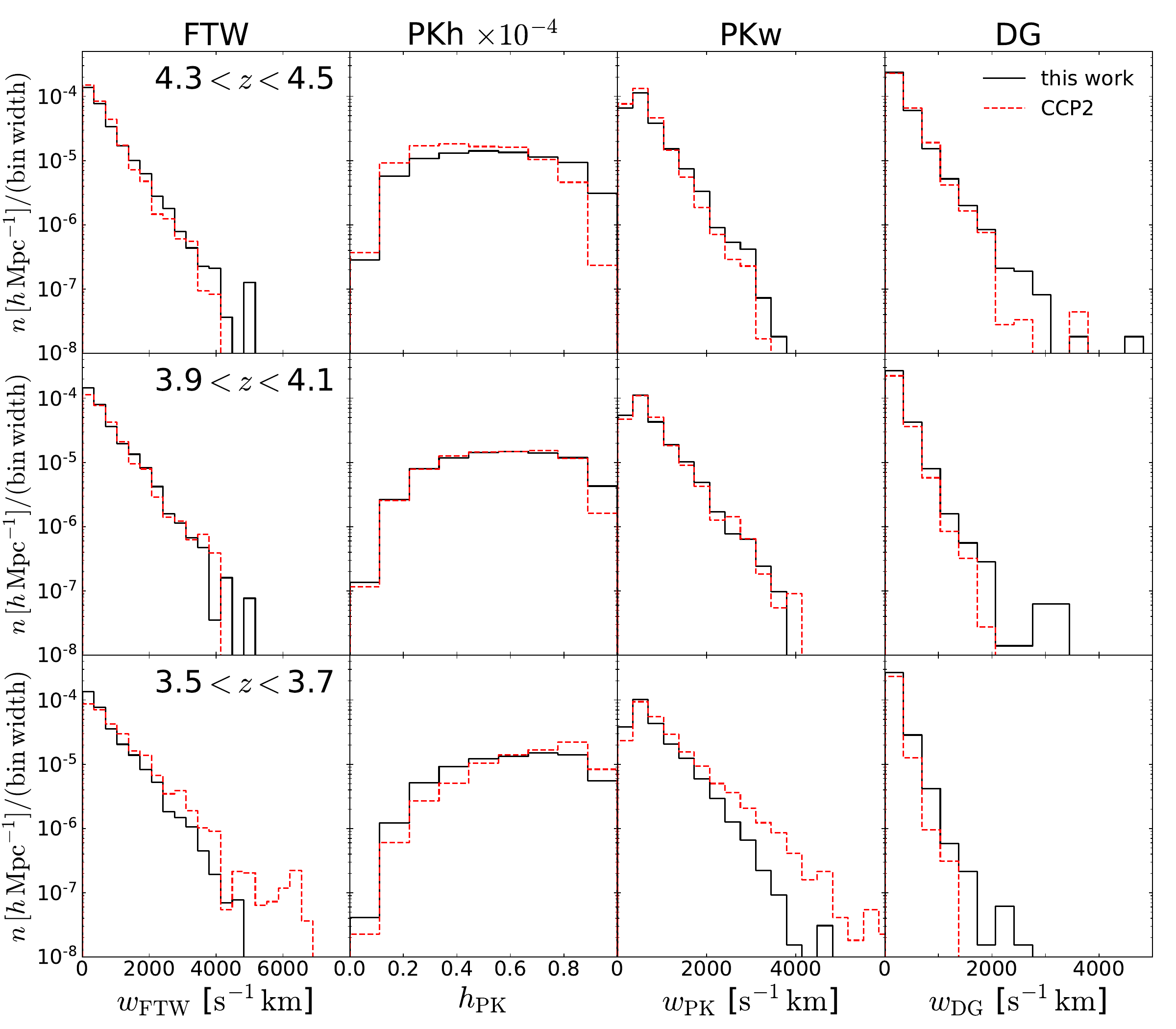}
\end{center}
\caption{Distribution of shape properties of transmission peaks and dark gaps in synthetic spectra of the \HI \Lya forest. Each row refers to a different redshift reported  in the leftmost 
panel. Each column shows the density distribution along synthetic sightlines of (from left to right) the width of flux transmission windows, the height (rescaled by a factor $10^{-4}$ for visual clarity) and width of transmission peaks and the size of dark gaps. A precise definition of these quantities is given in the text. Solid lines show the result for our simulations while dashed lines refer to the standard reionization history and employ the simulations of \CCPtwo.}
\label{fig:FTW+PKS+DG_H}
\end{figure*}

\subsection{Statistical properties of peaks and gaps}
\label{sec:peak_stat}

\begin{figure*}
\begin{center}
\includegraphics[width=\textwidth]{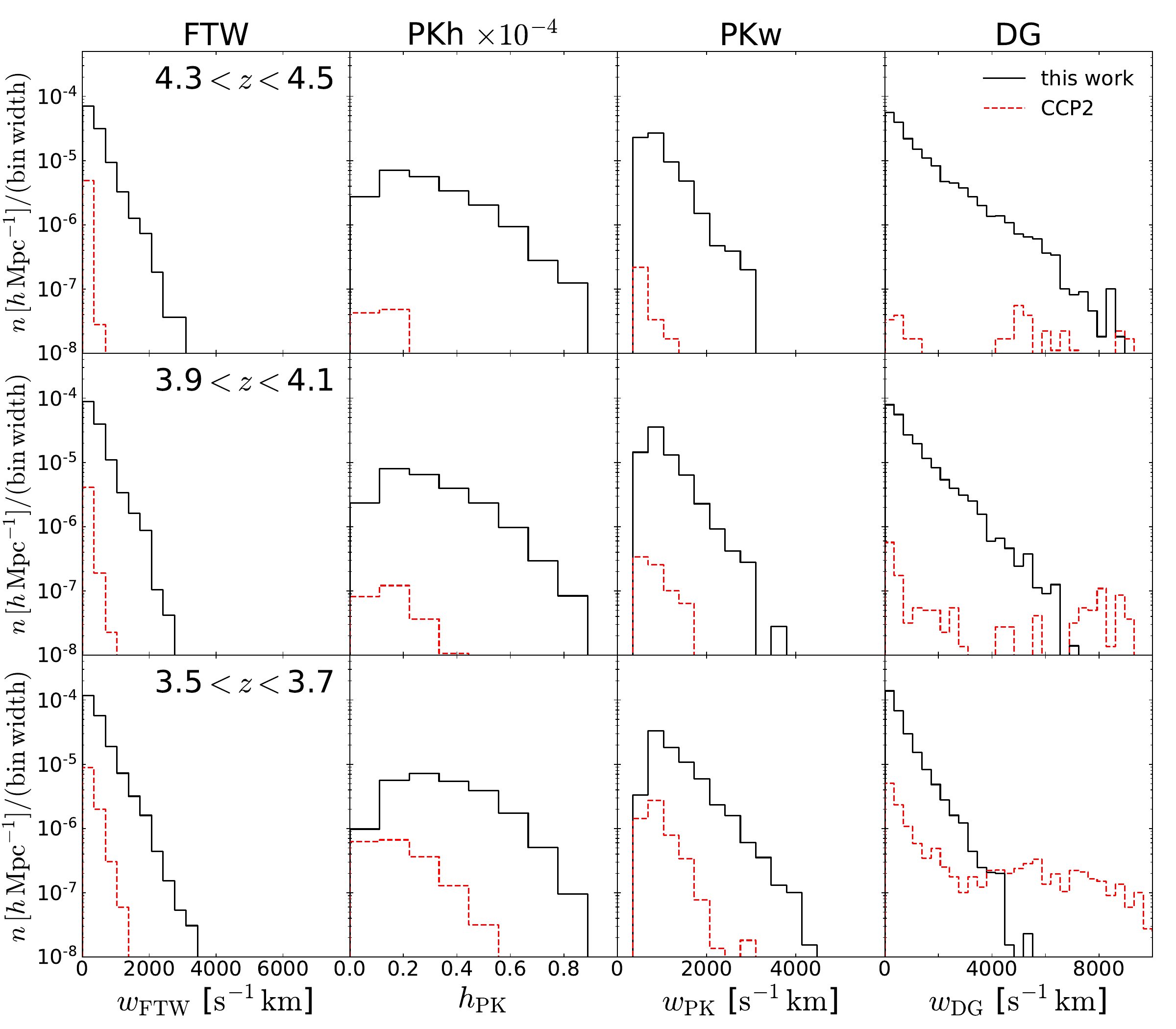}
\end{center}
\caption{As in Fig.~\ref{fig:FTW+PKS+DG_H} but for the \HeII \Lya forest. Notice the different horizontal scale in the rightmost column.}
\label{fig:FTW+PKS+DG_He}
\end{figure*}

In this Section and the next one, we use
our synthetic spectra to explore possible ways of gauging
the QSO contribution to cosmic reionization
with observational data. We proceed
by comparing the 
predictions of the QSO-dominated scenario with those of the
standard model.
For the latter, we use the results presented in \CCPtwo.
Our analysis does not involve any fine tuning of the model parameters 
\enrico{. Moreover, the statistics discussed here may not be fully converged at the resolution achieved by our simulation suite. As such, our study }
is qualitative in nature. Our intention is to
provide a guideline for future studies that will employ hybrid populations
of ionizing sources
in order to make more quantitative statements.

We first isolate specific features generated by \HI and \HeII \Lya absorption in QSO spectra and characterize them in terms of
four numbers.
Following \CCPone, we consider the width of flux-transmission windows (FTW) defined
as the (simply connected) regions where
the transmitted flux is everywhere greater than 20 per cent of the continuum level.
We also examine the length of dark gaps (DG) that are intended as the extended regions where the flux is everywhere below 10 per cent of the continuum level.
Finally, we look at transmission peaks
\citep{Gallerani+2008, Gnedin+2016} that we
define as the continuous regions 
where the flux always lies above 
a threshold value of 0.5 times the maximum
transmitted flux within the segment.
In particular, we record the width and the maximum height of the peaks (hereafter PKw and PKh, respectively).
For this analysis, in order to avoid classifying small local fluctuations as peaks, we add Gaussian noise with an rms value of $F_{\rm noise} = 0.05$ to the synthetic spectra and only consider peaks with a signal-to-noise ratio greater than 3.

Our results are shown in Fig.~\ref{fig:FTW+PKS+DG_H}
for the \HI \Lya forest 
and in Fig.~\ref{fig:FTW+PKS+DG_He} for the \HeII spectra.
In both cases, we compare the quasar-only scenario (solid lines) with
the standard one as computed in \CCPtwo (dashed lines) and focus on $z \lesssim 4.5$.
Shown are the density distributions of the different features per comoving pathlength in $\hMpc$.
For \HI, the two models give very similar results as hydrogen is highly ionized in the post-overlap phase and 
the transmissivity in the spectra is mostly dictated by the underlying density field. 
Small deviations are noticeable at late times because the models generate different intensities of the UV background. 
Overall, the distribution of DG widths ($w_{\rm DG}$) shifts towards shorter
values as the redshift decreases (see the rightmost column in Fig.~\ref{fig:FTW+PKS+DG_H}) as a consequence of the increasing ionization level of the IGM. 
Complementarily, the widths of FTWs ($w_{\rm FTW}$, leftmost column) and PKs ($w_{\rm PK}$, third column from the left) tend to increase. 
On the other hand, the distribution of peak heights ($h_{\rm PK}$, second column) 
hardly changes with time.

The two scenarios for the EoR, however, make very different predictions for the \HeII spectra.
Because of the late reionization of \HeII,
the standard model of reionization generates many less features than the QSO-dominated scenario.
Moreover, their distributions rapidly evolve with time thus showing the opposite trend as in the
QSO-dominated scenario where \HeII reionization takes place much earlier.
It is worth noticing that, in the standard case, the widths of FTWs and PKs are usually smaller than for \HI as a consequence of the reduced number of ionizing photons available. 
Complementarily, the width of DGs is larger.
In particular,
at $z\approx 3.6$, the standard scenario generates a prominent tail of very long DGs with 
$w_{\rm DG} \gtrsim 4500 \,\kms$ 
that is not present in the QSO-dominated case because of the higher number density of hard photons.

\begin{figure}
\begin{center}
\includegraphics[width=\columnwidth]{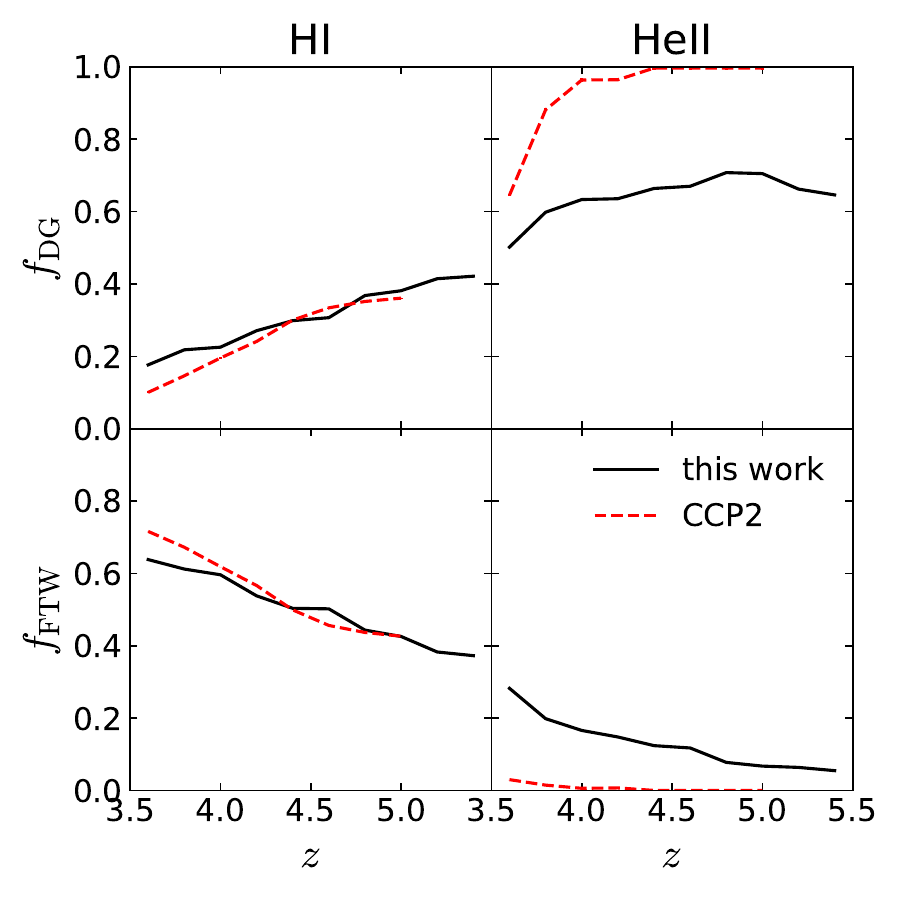}
\end{center}
\caption{Average fraction of a spectrum classified as DGs (top) or FTWs (bottom) as a function of redshift. The panels on the left-hand side refer to the \HI \Lya forest and those on the right-hand side to \HeII. The solid lines show the evolution in our set of simulations of the QSO-dominated scenario, while the dashed ones show the results of \CCPtwo for the standard model of the EoR. Statistical errors are always smaller than the line thickness.}
\label{fig:frac_FTW+DG}
\end{figure}

At first sight, it might be surprising that the number densities of both FTWs and DGs are small at high redshift. The reason is that our analysis
equally weights long and short features, \ie a completely absorbed spectrum will account for only one (long) DG, whilst a typical line of sight showing the \Lya forest 
will produce hundreds of DGs. 
In order to provide the missing information,
in Fig.~\ref{fig:frac_FTW+DG}, we plot the evolution of the mean fraction of the spectra which is assigned to DGs ($f_{\rm DG}$, top) or FTWs ($f_{\rm FTW}$, bottom). 
As expected, the portion of the spectra which is identified as FTWs increases with time while $f_{\rm DG}$ decreases. Once again, results for \HI are very similar in the two scenarios while they
strongly differ for \HeII, reflecting the different timing of helium reionization. In particular, in the standard model, there is basically no transmitted flux for $z\gtrsim 4.5$ while 30 to 40 per cent of the pixels at these
redshifts are not dark in the QSO-dominated scenario.

\subsection{The column-density ratio}
\label{sec:eta}

\begin{figure}
\begin{center}
\includegraphics[width=\columnwidth]{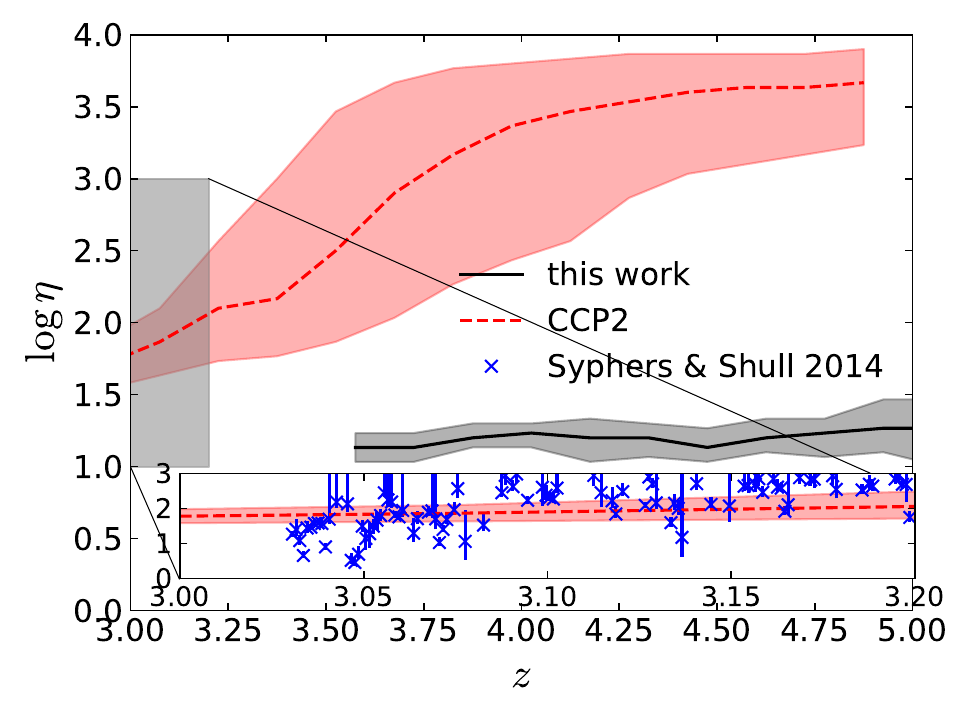}
\end{center}
\caption{Percentiles of the distribution of 
the column density ratio, $\eta$, as a function of redshift. The lines show the median value extracted from our simulations (solid) and from those in \CCPtwo (dashed) while the shaded regions enclose the central 68 per cent of the data.
The rectangular box on the left-hand side indicates the area containing the observational results by \citet{Syphers+2014} and is magnified in the inset for visual clarity.}
\label{fig:eta}
\end{figure}

An useful tool
to constrain the QSO contribution is 
the \HeII-to-\HI column-density ratio, $\eta$.
This quantity can be measured from QSO sightlines that are transparent in the Lyman continuum because no intervening Lyman limit systems block the UV part of the spectrum (hereafter \HeII QSOs).
For optically thin gas, the ratio encodes information on the spectral hardness of the ionizing radiation and thus on the relative contributions from hard (QSOs) and soft (stellar) sources.

In Fig.~\ref{fig:eta}, we show how
the probability density of $\eta$ 
evolves as a function of redshift
in our simulations and in those by \CCPtwo.
The solid lines show the median column-density ratio 
and the dashed lines enclose the central 68 per cent of the data. 
Results are very different in the standard and in the QSO-dominated scenarios due to the different timing of He reionization.
In the standard model, right after hydrogen reionization is completed, $\eta$ assumes values of several thousands (meaning that little or no flux is transmitted at the frequency of the \HeII Ly$\alpha$ transition) that tend to decrease with time and show a large scatter. In this case, $\eta$ 
fluctuations mainly trace the \HeII density
and the patchiness of \HeII-ionizing radiation.
Later on, when also \HeII is fully ionized, the $\eta$ distribution presents a broad peak at $\eta \sim 200$ \citep[see also][]{Meiksin+2012}.
On the other hand,
in the QSO-dominated model, \HI and \HeII are ionized in parallel and
their column-density ratio assumes significantly lower values, typically ranging between a few and a hundred at $z>6$.
As soon as the reionization of both 
species is completed, the PDF of $\eta$ rapidly relaxes to its final form which
sharply peaks at $\eta \sim 14$ and presents very little scatter.

Observations should be able to distinguish between
these very different evolutionary paths and final states.
Current statistical samples at $z \lesssim 2.7$ \citep{Shull+2010} provide a better match to
the $\eta$-distribution generated during the standard reionization history. Consistently, rare data at higher redshifts
present signs of incomplete \HeII reionization at $z\gtrsim 3$ \citep[\eg][]{Syphers+2014}.
It is thus plausible that 
collecting more \HeII-absorption spectra at $z\sim 3$ and contrasting them with custom-made numerical simulations might help us to precisely gauge the importance of the quasar contribution to the EoR.

\section{Discussion and Conclusions}
\label{sec:conclusions}
Determining the nature of the astrophysical sources of radiation that shaped the EoR continues to be an elusive goal. The tentative detection of a population of faint QSO candidates at high redshifts \citep{Giallongo+2015} led \MH to investigate a scenario in which the emission from active galactic nuclei dominates over the contribution of star-forming galaxies at all times. Their analytical calculations indicate that such a model is compatible with observations of the \HI volume fraction and with the optical depth to Thomson scattering of the cosmic microwave background.
In this paper, we have further scrutinized the QSO-dominated scenario by using more sophisticated tools.
We have run full hydrodynamical simulations and post-processed their output with a radiative-transfer code 
in order to compute the evolution of the temperature and ionization state of the IGM.
Subsequently, 
we assess the viability of the QSO-dominated scenario 
by producing synthetic absorption-line \HI and \HeII spectra from our simulations and 
comparing their properties with observations and with previous studies of the standard model for the EoR.

Our main results can be summarized as follows.
\begin{enumerate}
\item The \HII  and \HeIII volume fractions extracted from our simulations of the QSO-dominated scenario are consistent with most observational constraints and with the analytical model by \MH. 
Similarly, the Thomson optical depth of the IGM is in very good agreement with the latest 
measurements \citep{Planck2018cosmo}.

\item The striking feature that characterises the QSO-only scenario is that \HI, \HeI, and \HeII reionization take place nearly at the same time. In consequence, the IGM temperature at mean density shows a single peak at redshift $z \approx 5.5$ 
(where $T_{0, \rm peak} \approx 2\times10^4\,\rm K$). Compared with the bulk of the observed values, the model over predicts the IGM temperature at  $4<z<5$  and under predicts it at $2<z<3$. In particular, due to the early completion of \HeII reionization, the QSO-dominated model is inconsistent with the measurements that show a temperature peak at $z \approx 3$.
Our results are largely in agreement with the semi-analytical calculations presented by \citet{DAloisio+2017} although the IGM temperature peaks at an earlier time in our simulations.

\item Correspondingly, the  effective optical depth derived from our \HeII \Lya synthetic spectra is significantly too low at $3<z<4$ to reproduce the observational constraints.

\item The QSO-dominated model overpredicts opacity fluctuations in the \HI \Lya forest at all redshifts. When taken together, this feature, 
the under-estimation of the \HeII effective optical depth,
and the temperature evolution at odds with observations, constitute a (most likely irremediable) challenge for the scenario in which AGNs fully shape the EoR.

\item The redshift evolution of several features in the \HeII \Lya absorption spectra easily differentiates the QSO-dominated model for the EoR from the standard one and could be used to set tight constraints on the onset of \HeII reionization. Conversely, the properties of the \HI \Lya forest are very similar in the two scenarios.
Therefore, major progress in the field could be achieved by increasing the size of current samples of \HeII quasars and extending them to higher redshifts.

\item Although the QSO-dominated model is not able to fully reproduce the observed PDF of the \HI effective optical depth at $z\gtrsim 4.5$, it nicely generates very extended tails at high values that are not present
in the standard scenario where the reionization of \HI is much more spatially homogeneous and less patchy. This provides a hint that complementing the standard scenario with a sub-dominant population of high-$z$ QSOs might be key to reconcile the observed distribution of optical depths with the predictions from numerical simulations.
However, a `Goldilocks problem' emerges: the QSO contribution to the EoR needs to be highly fine tuned in order to match the distribution of optical depths without generating tension with other observables.

\end{enumerate}

In brief, our principal conclusion is that existing constraints on the IGM temperature and \HeII opacity rule out the QSO-dominated scenario we have investigated. There is a possible caveat, however.
Throughout the paper, we have assumed that all UV photons escape their sources independently of wavelength.
This is a common expectation motivated by the large luminosity of active galactic nuclei, 
although it has not been tested for the faint sources that generate most of the ionising photons at high redshift. 
By relaxing the hypothesis that $\fesc=1$ across all (relevant) wavelengths and
assuming that  $\fesc(\sim 912 \, \AA) > \fesc(\sim 228 \, \AA)$, it should be possible
to delay the onset of \HeII reionization and vastly improve the agreement with current observational constraints on the temperature of the IGM and the \HeII opacity.
This, however, will not modify much the PDF of the \HI optical depth which is mainly influenced by  
radiation close to the hydrogen ionization threshold. Nevertheless, this distribution is also sensitive
to sub-dominant contributions to the ionizing flux and dedicated simulations including also variable levels of stellar UV radiation need to be performed to address this issue in a more quantitative way.

Our results suggest that a rather extreme fine tuning of the escape fraction might be necessary to bring the QSO-dominated model for the EoR in agreement with existing observational data. Our analysis also reveals that developing a fully quantitative understanding of populations of sources that are active in the different phases of the EoR requires that future observational campaigns will collect many more \HeII QSO spectra so that to enable statistical studies of their characteristic features.

\section*{Acknowledgements}
We thank Emanuele Giallongo, Andrea Grazian, Frederick Davies and the anonymous referee for useful comments. This work is carried out within the Collaborative Research Centre 956 
\virg{The Conditions and Impact of Star Formation}, sub-project C4, 
funded by the Deutsche Forschungsgemeinschaft (DFG). 
We are thankful to the community developing and maintaining software 
packages extensively used in our work, namely: Matplotlib \citep{matplotlib},
NumPy \citep{numpy}, SciPy \citep{scipy}, yt \citep{yt}.

\bibliographystyle{mnras}
\bibliography{bibliography}

\bsp	
\label{lastpage}
\end{document}